\documentclass[rnote,structabstract]{aa}  
\usepackage{graphicx}
\usepackage{txfonts}
\usepackage{natbib}
\bibpunct{(}{)}{;}{a}{}{,} % to follow the A&A style
\begin{document}
   \title{The near-infrared counterpart of 4U 1636--53\thanks{Based on observations collected at the European Southern Observatory, Chile, under ESO Programme ID 085.D-0456(D).}}

   \author{D. M. Russell\inst{1,2},
           K. O'Brien\inst{3},
           T. Mu\~noz-Darias\inst{2}\fnmsep\inst{4},
           P. Casella\inst{4,5},
           P. Gandhi\inst{6}
          \and
          M. G. Revnivtsev\inst{7}
          }

   \institute{Astronomical Institute `Anton Pannekoek', University of Amsterdam, P.O. Box 94249, 1090 GE Amsterdam, the Netherlands
         \and
             Instituto de Astrof\'isica de Canarias (IAC), V\'ia L\'actea s/n, La Laguna E-38205, S/C de Tenerife, Spain\\
              \email{russell@iac.es}
         \and
             Department of Physics, University of California, Santa Barbara, California 93106, USA\\
             \email{kobrien@physics.ucsb.edu}
         \and
             School of Physics and Astronomy, University of Southampton, Southampton, Hampshire SO17 1BJ, UK\\
             \email{t.munoz-darias@soton.ac.uk}
         \and
             INAF - Osservatorio Astronomico di Roma, Via Frascati 33, I-00040 Monteporzio Catone (Roma), Italy\\
             \email{piergiorgio.casella@oa-roma.inaf.it}
         \and
             ISAS, Japan Aerospace Exploration Agency, 3-1-1 Yoshinodai, chuo-ku, Sagamihara, Kanagawa 229-8510, Japan\\
             \email{pgandhi@astro.isas.jaxa.jp}
         \and
             Space Research Institute, Russian Academy of Sciences, Profsoyuznaya 84/32, 117997 Moscow, Russia\\
             \email{revnivtsev@iki.rssi.ru}
             }

   \date{Received ?? ??, 2011; accepted ?? ??, 2011}

  \abstract
   {The optical counterpart of the neutron star X-ray binary and well known X-ray burster, 4U 1636--53 (= 4U 1636--536 = V801 Ara) has been well studied for three decades. However to date, no infrared studies have been reported.}
   {Our aims are to identify and investigate the near-infrared (NIR) counterpart of 4U 1636--53.}
   {We present deep, $K_{\rm S}$-band ($2.2 \mu \mathrm{m}$) imaging of the region of 4U 1636--53 taken with the Infrared Spectrometer And Array Camera (ISAAC) on the Very Large Telescope. Archival optical and UV data are used to infer the $0.2-2.2 \mu \mathrm{m}$ spectral energy distribution (SED).}
   {One star is located at coordinates $\alpha =16$:40:55.57, $\delta =-53$:45:05.2 (J2000; $1\sigma$ positional uncertainty of $\sim 0.3$ arcsec) which is consistent with the known optical position of 4U 1636--53; its magnitude is $K_{\rm S} = 16.14 \pm 0.12$. This star is also detected in the 2MASS survey in $J$-band and has a magnitude of $J = 16.65 \pm 0.22$. Under the assumption that the persistent emission is largely unvarying, the $0.4-2.2 \mu \mathrm{m}$ de-reddened SED can be described by a power law; $F_\nu \propto \nu^{1.5 \pm  0.3}$, with some possible curvature ($F_\nu \propto \nu^{\lesssim 1.5}$) at $0.2-0.4 \mu \mathrm{m}$. The SED can be approximated by a blackbody of temperature $\sim 27 000$ K. This is typical for an active low-mass X-ray binary, and the emission can be explained by the outer regions of a (likely irradiated) accretion disc. We therefore interpret this $K_{\rm S}$-band star as the NIR counterpart.}
   {}


\titlerunning{The NIR counterpart of 4U 1636--53}
\authorrunning{Russell et al.}

   \maketitle
%
%________________________________________________________________

\section{Introduction}

Low mass X-ray binaries (LMXBs) are interacting binaries where a low mass donor is transferring material onto a neutron star or a black hole. In order to transport the excess of angular momentum outwards, an accretion disc is formed. In the disc, the gravitational potential energy is transformed into mainly X-ray radiation and kinetic energy, and temperatures approach $\sim 10^8$ K. The mass transfer rate supplied by the donor star, $\dot{M}_2$, is driven by the binary/donor evolution and mass transfer rates $\dot{M}_2>\dot{M}_{\rm crit}\sim 10^{-9}M_{\odot} ~ \mathrm{kg} ~ \mathrm{yr}^{-1}$ (where $M_{\odot}$ is the mass of the Sun in kg) result in persistently bright X-ray sources \citep{kinget96}. There are $\sim 200$ such bright ($L_{\rm X} \simeq 10^{36} - 10^{38} \mathrm{erg} ~ \mathrm{s}^{-1}$) LMXBs in the Galaxy and most of them harbour neutron stars as implied by the detection of pulsations and X-ray bursts resulting from nuclear burning caused by the accumulation of Hydrogen and Helium on their surfaces. They show energy spectra dominated by emission from their irradiated accretion discs which also dominate at optical wavelengths \citep[e.g.,][]{vanpet95}.

Even in the near-infrared (NIR), where the companion star could have an important contribution, the disc emission seems dominant in bright systems. For instance, in the prototypical persistent neutron star, Sco X--1, which given its relatively long orbital period ($\sim 19$ h) should have a large, evolved companion star, no spectral feature from the donor has been detected in the NIR \citep{bandet97}. In most persistent neutron star X-ray binaries, spectral and temporal studies have favoured an X-ray heated accretion disc as the origin of the NIR emission \citep[e.g., 4U 1705--440;][]{homaet09}. Compact jets producing synchrotron emission typically dominate the radio emission \citep{miglfe06} and their spectra can extend to higher frequencies. In the NIR, high amplitude flares from GX 17+2 \citep{callet02}, a synchrotron spectrum in 4U 0614+09 \citep{miglet10} and variable linear polarization in Sco X--1 \citep{russfe08} have suggested a strong infrared jet contribution in these persistent neutron star X-ray binaries.

One of the classical neutron star systems is 4U 1636--53 (= 4U 1636--536 = V801 Ara). It is an X-ray burster, which has been extensively studied in X-ray and optical regimes for more than three decades \citep[e.g.,][]{pedeet82}. It has a 3.8 h orbital period \citep{vanpet90,gileet02} pointing to a relatively faint, late type companion star. The spectrum of the system from X-ray to optical wavelengths seems to be dominated by the emission of its bright accretion disc, the companion star only being detected by using emission lines arising from reprocessing of the strong X-ray emission ($L_X \sim 10^{37 - 38} \mathrm{erg} ~ \mathrm{s}^{-1}$) in its inner hemisphere \citep{casaet06}.

To date, no detections of 4U 1636--53 have been reported at wavelengths longer than the optical regime. At radio frequencies, upper limits of both the persistent and burst fluxes were presented in \citeauthor{thomet79} (\citeyear{thomet79}). Here we present the first detection of the near-infrared counterpart of 4U 1636--53. Together with available optical and UV data, we construct the intrinsic 0.2--2.2 $\mu \mathrm{m}$ spectral energy distribution (SED).
%__________________________________________________________________

   \begin{figure}
   \centering
   \includegraphics[width=8.5cm,angle=0]{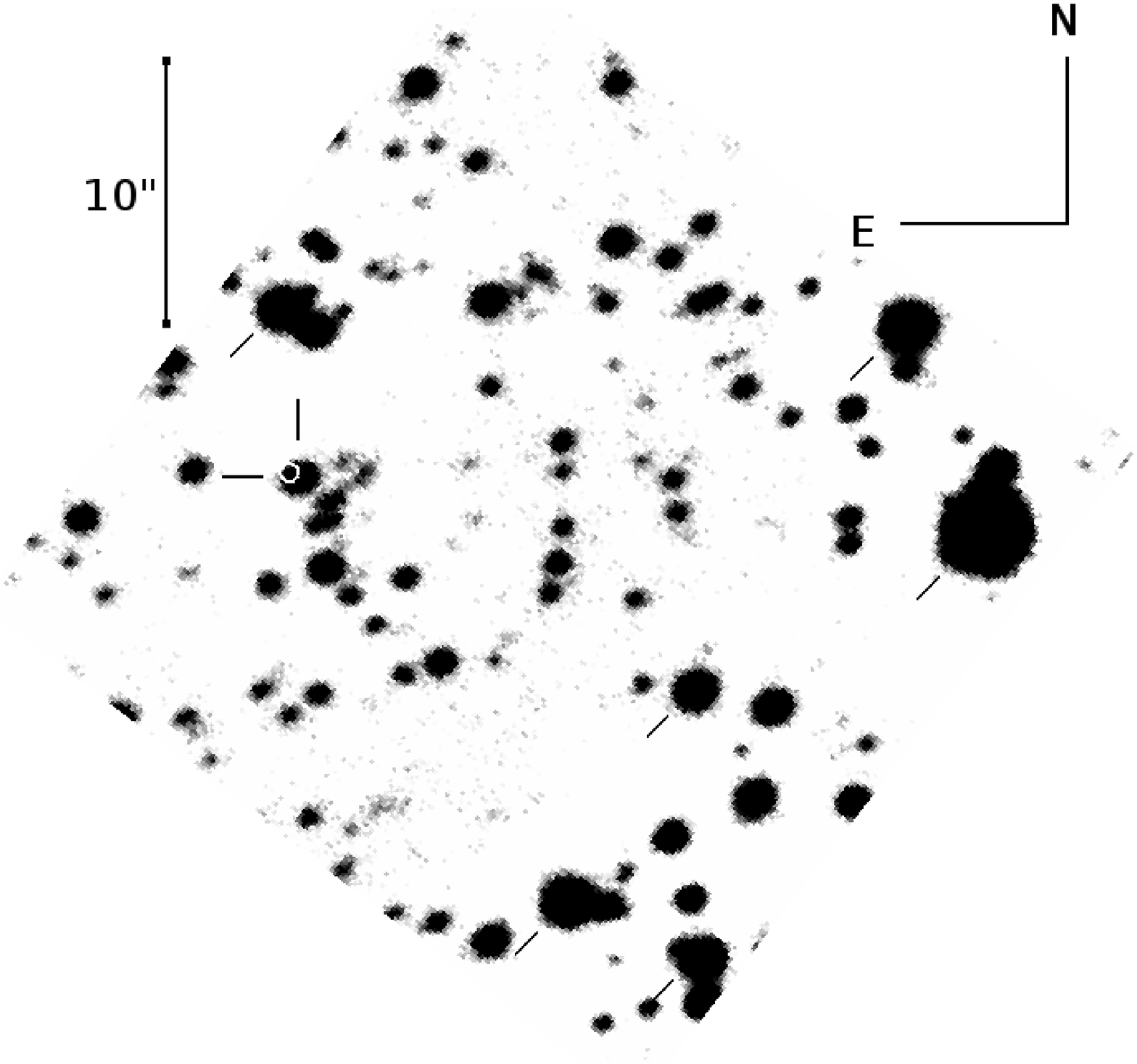}
   \caption{VLT / ISAAC deep, high resolution $K_{\rm S}$-band finding chart for 4U 1636--53. The optical position (USNO-B1 0.339\arcsec\ error circle) is indicated by a white circle and the infrared (ISAAC) position is shown by two black lines. The two positions agree within $1\sigma$ errors. The six 2MASS stars are indicated by diagonal lines.}
              \label{finding1}%
    \end{figure}

\section{Very Large Telescope observations}

The data were acquired with the Infrared Spectrometer And Array Camera (ISAAC) on the European Southern Observatory (ESO) 8-m class Very Large Telescope UT3 (Melipal) on 2010-05-21 02:23 -- 03:17 UT (MJD $55337.118 \pm 0.019$) under ESO Programme ID 085.D-0456(D). Twenty-four data cubes, each of exposure time 132.5 s (comprising $530 \times 0.25$ s individual exposures) were obtained of the region of 4U 1636--53 in high time resolution (FastJitter) mode using the $K_{\rm S}$-band filter centred at $2.16 \mu \mathrm{m}$. The brightest object in the field has $< 1000$ counts in each 0.25 s exposure, so non-linearities are not a concern. A small (up to 6\arcsec) telescope offset was applied between cubes. Average images were made of each cube, removing the sky generated from a median of the surrounding averaged cubes. The field of view of each ISAAC image is $40\arcsec \times 40\arcsec$ and the pixel scale is $0.1478\arcsec\ \mathrm{pixel}^{-1}$. The 24 average images were then aligned and stacked using \small IRAF\normalsize\ to produce a deep image. The total on-source exposure time of the combined image is 3180 s. The $K_{\rm S}$-band seeing as measured from the combined image was 0.6 arcsec and conditions were clear. The camera was rotated at an angle of $127^{\circ}$ to maximise the number of reference stars on the array. The combined image is presented in Fig. \ref{finding1} and can be used as a deep, high resolution $K_{\rm S}$-band finding chart.

\section{Source identification}

Within the combined image, six stars listed in the Two Micron All Sky Survey \citep[2MASS;][]{skruet06} are detected, which were used to calculate the world coordinate system and achieve $K_{\rm S}$-band flux calibration. \small DAOPHOT\normalsize\ in \small IRAF\normalsize\ is used to perform point-spread-function (PSF) photometry on the combined image. All comparison stars were successfully fitted by a single PSF consistent with the image PSF; no blends were identified. There is one star consistent with the best optical position of 4U 1636--53, as listed in the USNO-B1 digital sky survey ($\alpha =16$:40:55.61, $\delta =-53$:45:05.1; J2000, with an error circle of 0.339\arcsec), shown as a white circle in Fig. \ref{finding1}. We derive a position of $\alpha =16$:40:55.57, $\delta =-53$:45:05.2 (J2000) for the infrared counterpart of 4U 1636--53, with a $1\sigma$ positional uncertainty of $\sim 0.3$ arcsec. Other optical coordinates of 4U 1636--53 in the literature \citep{jernet77,liuet01,samuet03} are also consistent with the USNO-B1 and ISAAC positions. No precise X-ray coordinates of 4U 1636--53 from Chandra or Swift observations exist in the literature.

Some faint, blended stars lie just 1.5\arcsec\ from 4U 1636--53 in the ISAAC image, but none are consistent with the optical position. The above source is the only candidate counterpart in our $K_{\rm S}$-band image. All stars within a $3\arcsec$ radius of the LMXB are $\geq 1.2$ magnitudes fainter in $K_{\rm S}$-band than this proposed counterpart. No faint blended stars are visible after subtracting the PSF of the counterpart.

From the known magnitudes of the 2MASS stars within the field, we measure the magnitude of 4U 1636--53 during the time of our observations to be $K_{\rm S} = 16.14 \pm 0.12$. The errors are dominated by systematic errors derived from the six 2MASS star magnitudes. 4U 1636--53 is detected with a signal-to-noise ratio (S/N) of 145. The limiting magnitude in the image is $K_{\rm S} \sim 19.3$ (i.e. stars brighter than this are detected at the $3\sigma$ level).

The counterpart we have found is not listed in the 2MASS point source catalogue. On inspection of the 2MASS images, the counterpart is not visible in $K$ and $H$-bands, but appears as a faint source in $J$-band. Performing PSF photometry on the $J$-band 2MASS field, we obtain a magnitude of $J = 16.65 \pm 0.22$ for 4U 1636--53; the S/N is 7. The aforementioned faint stars within $\sim 2\arcsec$ of the source may contribute up to $\sim 10 - 30$\% of the flux measured, although the PSF fitting method should have removed the majority of this excess flux (which is offset from the central PSF position). This 2MASS observation was made on 1999-06-18 04:27 UT. In the 2MASS $H$-band image, the upper limit to the magnitude of 4U 1636--53 is $H > 15.76$. The new VLT and 2MASS detections of 4U 1636--53 are listed in Table 1.

   \begin{figure}
   \centering
   \includegraphics[width=8.5cm,angle=0]{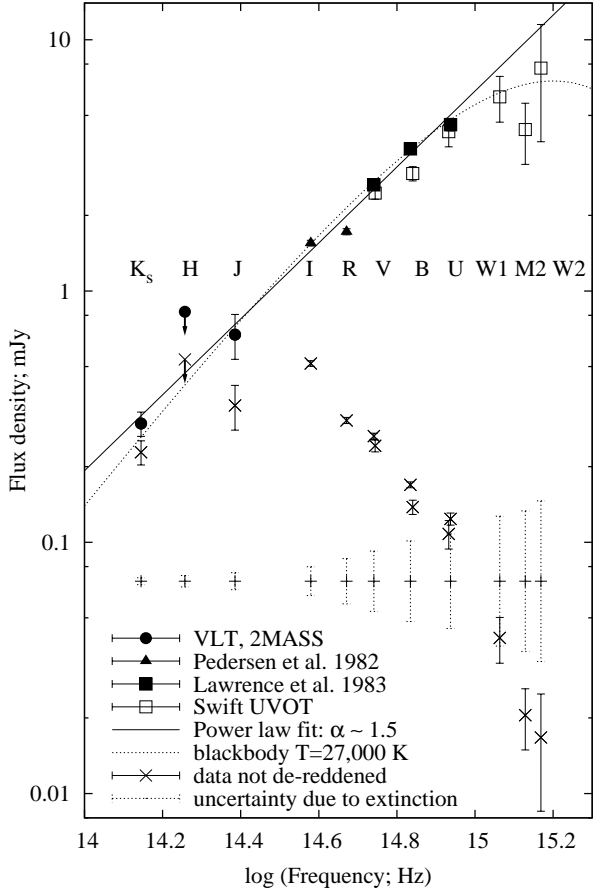}
   \caption{Intrinsic (de-reddened) UV--optical--NIR spectral energy distribution of 4U 1636--53 (top) and observed (reddened) fluxes (lower, crosses). The systematic errors due to uncertainties in the extinction are demonstrated by the dotted error bars.}
              \label{sed}%
    \end{figure}

\begin{table}
\begin{center}
\caption{New NIR -- optical -- UV magnitudes of 4U 1636--53. The \textit{Swift} UVOT photometry is taken between 2005-02-13 and 2010-06-10.}
\begin{tabular}{lllllllll}
\hline
Filter&$\lambda ^1$&S/N   &\multicolumn{3}{c}{---------- Magnitude ----------}&Days$^2$\\
\medskip
      &($\AA$)     &      &mean &brightest&faintest &    \\
\hline
$K_{\rm S}$&21590  &145   &16.14&16.14(12)&16.14(12)&--\\
J     &12350       &7     &16.65&16.65(22)&16.65(22)&--\\
$v$   &5468        &28    &17.94&17.94(6) &17.94(6) &1	 \\
$b$   &4392        &21    &18.67&18.67(8) &18.67(8) &1	 \\
$u$   &3465        &27--43&17.82&17.63(7) &17.95(8) &5	 \\
$uw1$ &2600        &10--24&18.35&18.16(9) &18.67(15)&4	 \\
$um2$ &2246        & 7--15&18.95&18.61(13)&19.31(20)&5	 \\
$uw2$ &1928        & 7--34&19.20&18.53(10)&19.61(21)&4	 \\
\hline
\end{tabular}
\end{center}
\small
\medskip
$^1$The central wavelength of the filter in Angstroms; $^2$The number of dates the source was observed with UVOT.
\normalsize
\end{table}

\section{Swift UVOT observations}

Optical--ultraviolet observations of 4U 1636--53 were made with the UltraViolet/Optical Telescope \citep[UVOT;][]{romiet05} on board the \textit{Swift} satellite and the data are publicly available. UVOT observed 4U 1636--53 on 11 dates between 2005-02-13 and 2010-06-10. On most dates, one or two of the six UVOT filters were used. The image data of each filter on each date were summed using {\tt uvotimsum}. Photometry of the source in individual sequences was derived with {\tt uvotsource}. 4U 1636--53 is clearly detected at the position derived above, in all six filters ($5500-1900\AA$). An extraction region of radius 3\arcsec\ centred at the ISAAC-derived position was adopted, which excludes a neighbouring star 6\arcsec\ away from 4U 1636--53; no stars appear to contaminate the flux of 4U 1636--53 within the extraction radius. The mean magnitudes and range in each filter are given in Table 1.

\section{Intrinsic spectral energy distribution}

We de-redden the $K_{\rm S}$, $J$-band and UVOT magnitudes using the known interstellar extinction towards 4U 1636--53; $A_{\rm V} = 2.5 \pm 0.3$ mag \citep[derived from optical data;][]{lawret83} applying the extinction law of \cite{cardet89}. We also take optical magnitudes from \cite{lawret83} and \cite{pedeet82}; $U = 17.92 \pm 0.06$, $B = 18.47 \pm 0.03$, $V = 17.89 \pm 0.03$, $R = 17.46 \pm 0.03$, $I = 16.77 \pm 0.03$, and de-redden them in the same manner. Using the de-reddened fluxes \citep[adopting the standard conversion zeropoints for optical Johnson filters, and 2MASS IR filters from the Explanationary Supplement of][]{skruet06} we construct the intrinsic NIR--optical--UV SED of the persistent (non-burst) emission from 4U 1636--53.

The SED, which spans one order of magnitude in frequency, is presented in Fig. \ref{sed}. The observed, reddened fluxes are plotted in addition to the de-reddened data. The optical ($I$ to $U$-band) SED from \cite{lawret83} and \cite{pedeet82}, which has small errors, has a spectral index of $\alpha = 1.5 \pm 0.4$, where $F_\nu \propto \nu^{\alpha}$. The error on $\alpha$ is dominated by the uncertainty in the extinction. This spectral index is typical for active low-mass X-ray binaries \citep[both neutron star and black hole systems; e.g.,][]{hyne05,russet07} and is consistent with the outer regions of a blue accretion disc. \citeauthor{shihet11} (\citeyear{shihet11}) showed that the optical emission is correlated with the soft X-ray flux; the optical light is dominated by the irradiated disc (X-ray reprocessing on the disc surface).

The $K_{\rm S}$, $H$-band and $J$-band magnitudes correspond to de-reddened flux densities of $F_{\nu,K_{\rm S}} = 0.30 \pm 0.03$ mJy, $F_{\nu,H} < 0.83$ mJy and $F_{\nu,J} = 0.67 \pm 0.14$ mJy, respectively. The optical--infrared ($K_{\rm S}$ to $U$-band) SED (neglecting the UVOT data) can be fitted with a power law with spectral index $\alpha = 1.5 \pm 0.3$ (solid line in Fig. \ref{sed}). This is similar to the optical spectral index, and the $K_{\rm S}$-band and $J$-band fluxes are consistent with an extrapolation of the accretion disc spectrum. It is unlikely that IR emission lines \citep[e.g.][]{bandet97} could contribute a significant fraction of the observed IR fluxes. There is some possible curvature in the spectrum at the higher frequencies, apparent in the UVOT data, and the whole SED can be approximated by a blackbody at a temperature of $\sim 27~000$ K (dotted line in Fig. \ref{sed}). The curvature implies the blackbody of the irradiated disc spectrum may peak at around 1900 $\AA$, similar to some other LMXBs \citep{hyne05}. However, the curvature (and even more so the spectral index of the power law) is sensitive to the uncertainty in the extinction, and the optical counterpart varies by $\sim 0.6$ mag amplitude (a factor of $\sim 1.7$ in flux) over weeks to months \citep[e.g.,][]{shihet11}. We detect variability in our UVOT data over several years (Table 1). Quasi-simultaneous optical--infrared data are required to accurately measure the spectral index, and uncertainties will be decreased once the extinction towards the source is measured more accurately.

There is no evidence for the companion star, or synchrotron emission from the jet (if the system has a jet) to dominate the infrared flux. Both of these components are expected to be redder than the irradiated disc component, and would produce an infrared excess above the disc spectrum. Since the spectrum between $J$ and $K_{\rm S}$-bands is blue ($\alpha \sim 1.5$), these components make a negligible contribution to $J$-band. In the most extreme scenario, the disc could have a steep spectrum, $\alpha = 2.0$ between $J$ and $K_{\rm S}$-band, and therefore the disc could be as faint as 0.18 mJy in $K_{\rm S}$-band ($0.12 \pm 0.03$ mJy fainter than the observed flux). If this were the case, the jet or companion could contribute up to 46\% of the $K_{\rm S}$-band flux. These calculations again neglect possible variability, since the NIR observations were made on different dates. Simultaneous optical--infrared observations could constrain the level of star or jet emission more precisely.

\section{Conclusions}

We have identified the near-infrared counterpart of the neutron star burster LMXB 4U 1636--53. Its magnitudes on the dates observed are $K_{\rm S} = 16.14 \pm 0.12$ (VLT / ISAAC), $J = 16.65 \pm 0.22$ (2MASS). The intrinsic infrared--optical--UV spectrum of the persistent (non-burst) emission is consistent with a blackbody, likely from the irradiated surface of the accretion disc. We find no evidence for emission from other components that may be expected to contribute, such as the donor star or synchrotron emission from a jet, although simultaneous optical and infrared data are needed to constrain these contributions further.

\begin{acknowledgements}
This research was partly supported by a Netherlands Organisation for Scientific Research (NWO) Veni Fellowship and partly by a Marie Curie Intra European Fellowship within the 7th European Community Framework Programme under contract no. IEF 274805. The research leading to these results has received funding from the European Communitys Seventh Framework Programme (FP7/2007-2013) under grant agreement number ITN 215212 -Black Hole Universe-.  Partially funded by the Spanish MEC under the Consolider-Ingenio 2010 Program grant CSD2006-00070: First Science with the GTC (http://www.iac.es/consolider-ingenio-gtc/). TMD acknowledges support by ERC advanced investigator grant 267697-4PI-SKY. PC acknowledges funding via a EU Marie Curie Intra-European Fellowship under contract no. 2009-237722. MR is supported by grants MD-1832.2011.2, RFBR 10-02-00492 and by Dynasty foundation. This publication makes use of data products from the Two Micron All Sky Survey, which is a joint project of the University of Massachusetts and the Infrared Processing and Analysis Center/California Institute of Technology, funded by the National Aeronautics and Space Administration and the National Science Foundation.
\end{acknowledgements}

\bibliographystyle{aa} % style aa.bst

\end{document}